\documentclass[10pt]{iopart}
\usepackage{amssymb}
\usepackage{graphicx}
\eqnobysec

\newcommand{\beq}{\begin{equation}}
\newcommand{\beqa}{\begin{eqnarray}}
\newcommand{\eeq}{\end{equation}}
\newcommand{\eeqa}{\end{eqnarray}}
\renewcommand{\b}{^{(b)}}
\newcommand{\dd}{{\rm d}}
\newcommand{\eps}{\varepsilon}
\newcommand{\frad}[2]{{\displaystyle{\displaystyle#1\over\displaystyle#2}}}
\newcommand{\mean}[1]{\left\langle#1\right\rangle}
\newcommand{\stir}[2]{\left[#1\atop#2\right]}
\renewcommand{\u}[1]{{\underline{#1}}}
\newcommand{\Int}[1]{{\lfloor#1\rfloor}}
\renewcommand{\P}{{\cal P}}

\begin{document}

\title{Coverage fluctuations in theater models}

\author{P L Krapivsky$^{1,2}$ and J M Luck$^2$}

\address{$^1$ Department of Physics, Boston University, Boston, MA 02215, USA}

\address{$^2$ Institut de Physique Th\'eorique, Universit\'e Paris-Saclay,
CEA and CNRS, 91191 Gif-sur-Yvette, France}

\begin{abstract}
We introduce the theater model,
which is the simplest variant of directed random sequential adsorption
in one dimension with point source and steric interactions.
Particles enter sequentially an initially empty row of $L$ sites and adsorb
irreversibly at randomly chosen places.
If two particles occupy adjacent sites,
they prevent further particles from passing them.
A jammed configuration without available empty sites is eventually reached.
More generally, we investigate the class of models parametrized by $b$,
the number of consecutive particles needed to form a blockage.
We show analytically that the occupations of different sites in jammed configurations
exhibit long-range correlations obeying scaling laws, for all integers $b\ge2$,
so that the total number of particles grows as a subextensive power of $L$,
with exponent $(b-1)/b$, and keeps fluctuating even for very large systems.
The exactly known relative number variance measuring this lack of
self-averaging is maximal for the theater model {\it stricto sensu} ($b=2$).
In the special case where $b=1$, so that each adsorbed particle is a blockage,
the model can be mapped onto the statistics of records in sequences of random
variables and of cycles in random permutations.
A two-sided variant of the model is also considered.
In both situations the number of particles grows only logarithmically with $L$,
and it is self-averaging.
\end{abstract}

\eads{\mailto{pkrapivsky@gmail.com},\mailto{jean-marc.luck@ipht.fr}}

\maketitle

\section{Introduction}

Random sequential adsorption (RSA)
is the simplest of all models describing totally irreversible
dynamics~\cite{evans,talbot,KRB}.
It is relevant to a wealth of physical situations
ranging from chemical reactions on polymers to crystal growth and glass formation.
Particles are adsorbed irreversibly on a homogeneous substrate,
subject to some local rule such as nearest neighbor avoidance on a lattice.
Historical examples include dimer deposition on an infinite chain,
studied by Flory in 1939~\cite{flory},
and the car parking problem on a continuous line,
solved in 1958 by R\'enyi~\cite{re1,re2}.
The dynamics eventually stops when the system reaches a jammed (i.e., fully blocked)
configuration, where no further particle can be added.
The quantity of main interest is the limit density (or coverage) of the system,
i.e., the fraction of space occupied by particles in jammed configurations.
For standard RSA on a homogeneous substrate,
this quantity is self-averaging in the thermodynamic limit,
in the sense that relative coverage fluctuations become negligible for larger
and larger systems.
The final coverage however depends on details such as e.g.~the initial coverage,
whenever the latter is non-zero~\cite{BK,DS}.

In this paper we introduce and study the following {\it theater model},
which is the simplest directed and inhomogeneous avatar of RSA in one dimension
with point source and steric interactions.
To our knowledge, this model is novel, in spite of its simplicity.
A row of $L$ initially empty sites
is occupied by particles according to the following rules:

\begin{itemize}

\item Particles enter the system one by one from the left.

\item Each particle randomly selects an available empty site and occupies it
forever.

\item If two particles occupy adjacent sites,
they prevent further particles from passing them.
Only empty sites to the left of the blocking pair remain available.

\end{itemize}

The system can be viewed as a row of seats in a theater,
where latecomers are ready to disturb singles
in order to access further available seats,
but unwilling to disturb sitting couples.
A model in the same vein has already been considered in~\cite{greek},
albeit without steric interactions,
which are the key novel ingredient of the present theater model.
A possible microscopic realization of our model at the molecular scale
is that of a narrow channel, open at one end, such as e.g.~in a zeolite,
where molecules may enter and adsorb anywhere,
and pairs of nearby adsorbed molecules rearrange their conformation
and thus hinder the passage of subsequent ones.
As it turns out,
mechanisms of this kind have been suggested recently in the case of
benzene~\cite{chem1} and of other aromatic molecules~\cite{chem2}.
Other directed variants of RSA in one dimension
have been investigated in the framework of polymer translocation through a pore
in a membrane~\cite{poly1,poly2}.
Finally, RSA on more general inhomogeneous substrates has also been studied
by means of density functional theory~\cite{DFT}.

Here
\beq
\label{example}
\rightarrow\,\square\,\square\,\bullet\,\square\,\square\,\bullet\,
\square\,\square\,\square\,\u{\bullet\,\bullet}\,\circ\,\circ\,\circ\,\bullet
\eeq
is an example of a partly filled row of length $L=15$ with
five occupied sites (denoted by $\bullet$), one blockage (underlined),
seven available empty sites (denoted by $\square$)
and three blocked empty sites (denoted by $\circ$).

A jammed configuration without available empty sites is eventually reached.
Here
\beq
\rightarrow\,\bullet\,\bullet\,\bullet\,\circ\,\circ\,\bullet\,\bullet\,\circ\,
\circ\,\bullet\,\bullet\,\circ\,\circ\,\circ\,\bullet
\label{jammed}
\eeq
is an example of a jammed configuration reached from~(\ref{example}) by
filling three more sites.
In each jammed configuration the first two sites are occupied.
Every configuration whose first two sites are occupied
may actually be reached as a jammed configuration of the model, and so
\beq
\rightarrow\,\bullet\,\bullet\,\circ\,\circ\,\circ\,\circ\,\circ\,\circ\,
\circ\,\circ\,\circ\,\circ\,\circ\,\circ\;\circ
\label{min:jammed}
\eeq
and
\beq
\rightarrow\,\bullet\,\bullet\,\bullet\,\bullet\,\bullet\,\bullet\,\bullet\,
\bullet\,\bullet\,\bullet\,\bullet\,\bullet\,\bullet\,\bullet\;\bullet
\label{max:jammed}
\eeq
are respectively the least dense and the densest of all jammed configurations.

The model can be extended by requiring that any number $b$
of adjacent occupied sites are needed to constitute a blockage.
The only parameters of the extended model are the integers $b$ and $L$.
The theater model introduced above corresponds to $b=2$.
The particles experience steric interactions for all $b\ge2$.
The situation where $b=1$ is already non-trivial,
in spite of the absence of interactions.
It is a special case of the model investigated in~\cite{greek}
(see section~\ref{secun}).

The following general properties of the model and of its jammed configurations
hold for all integers $b\ge1$ and will be instrumental in the sequel.
Throughout the following, sites are numbered as $n=1,\dots,L$, from left to right.

\begin{itemize}

\item
Interactions are fully directed.
The occupation of any site $n$ is only affected by the sites to its left ($m=1,\dots,n-1$).
As a consequence, local properties of jammed configurations
do not depend on the system size~$L$, provided the latter is large enough.
For instance, the single-site occupation probability $p_n$ is independent of~$L$
as soon as $L\ge n$.
This absence of finite-size effects is a common property of one-dimensional systems
enjoying {\it spatial causality}, in this very sense that interactions are fully directed.
Earlier examples of processes exhibiting this feature
include a model for the orientational dynamics of a column of grains~\cite{ML1,SLM},
asymmetric annihilation processes~\cite{AM},
and a spin chain endowed with disordered asymmetric dynamics~\cite{AA}.

\item
The dynamics is fully irreversible.
Therefore, only the first attempt at filling a given site may be successful.
If the site is available and empty, the particle occupies it forever.
If it is already either occupied or blocked, it will remain so forever.
In neither case can a second visit be a success.
This property implies that it is more convenient to describe a history of the
system in terms of the first visits to its sites,
rather than in terms of individual incoming particles.
In particular, the jammed configuration reached by the process
only depends on the ordering in time of the first visits to the~$L$ sites.

\item
In this work, we are only interested in the statistics of jammed
configurations, and not in time-dependent quantities.
The times at which the sites are visited first can therefore be modelled at our discretion.
Here we choose a Poisson process, where the time $t_n$ of the first visit to
site $n$, i.e., of the first ---and only possibly successful--- attempt at filling it,
is modelled as an exponential random variable with density $\e^{-t}$.
Equivalently, the {\it height variables}
\beq
x_n=\e^{-t_n}
\label{xdef}
\eeq
are independent uniform variables on [0, 1].
These variables can be viewed pictorially
as a static random height profile over the system.
Within this level of description,
a history of the process corresponds to a uniform draw of the~$L$ height variables.
The ordering of those variables entirely determines the jammed configuration.
These orderings are in one-to-one correspondence with permutations of $L$ objects.
This correspondence has been instrumental,
e.g.~in the study of patterns of rises and falls in random sequences
(see~\cite{bou} and the references therein).

\end{itemize}

The focus of this work is on the statistics of jammed configurations.
We shall be mostly interested in the distribution $P(N,L)$
of the number $N$ of particles in those configurations in a system of size $L$.
Other quantities of interest are the probabilities~$p_n$ that site $n$ is occupied,
$p_{m,n}$ that sites $m$ and $n$ are simultaneously occupied, and so on,
These probabilities are independent of the system size $L$ as soon as $L$ is large enough.
However, at variance with standard RSA, configurations are inhomogeneous.
The occupation probability $p_n$ slowly falls off to zero
as a negative power of the distance~$n$ to the entry point,
whereas higher-order occupation probabilities exhibit non-trivial long-range correlations
for all $b\ge2$.
Both classes of observables introduced above are related to each other.
For instance, the mean particle number in a system of size $L$ reads
\beq
\mean{N}=\sum_{N=1}^LP(N,L)=\sum_{n=1}^Lp_n.
\label{ave}
\eeq

The setup of this paper is as follows.
Section~\ref{secun} is devoted to the case where $b=1$,
already considered in~\cite{greek}.
This situation, where every occupied site is a blockage,
is solvable by means of an exact mapping between
jammed configurations and sequences of records.
The occupation probability is exactly $p_n=1/n$,
whereas the number of particles is self-averaging and grows as $N\approx\ln L$.
The two-sided variant of the model,
where particles may enter from either end of the system,
is also studied by analytical means.
The number of particles is again self-averaging and grows as $N\approx2\ln L$.
Section~\ref{secde} contains a detailed investigation of the theater model
{\it stricto sensu} ($b=2$).
We first derive combinatorial results for finite systems
(section~\ref{secdeex}),
obtaining exact rational expressions for the single-site occupation probability,
the mean number of particles, the pair occupation probability
and the probabilities of the least dense and densest configurations.
We then derive asymptotic results on large systems (section~\ref{secdeasy}).
The single-site and pair occupation probabilities respectively fall off as
$p_n\approx\sqrt{\pi}/(2\sqrt{n})$ and $p_{n,n+1}\approx1/n$,
whereas higher-order joint occupation probabilities exhibit long-range
correlations with power-law scaling.
The distribution of the number of particles, too, obeys an asymptotic scaling law.
The rescaled variable
\beq
\nu=\frac{N}{\sqrt{L}}
\eeq
has a non-trivial limit law $\P(\nu)$, demonstrating that
the number of particles keeps fluctuating and does not become self-averaging in
the thermodynamic limit.
The first three moments of the latter limit law are determined explicitly,
as well as its decay at small and large $\nu$.
The main outcomes of this analysis
are then extended in section~\ref{sechi} to all integers $b\ge2$.
We derive first exact expressions
of the probabilities of the least dense and densest configurations
on finite systems (section~\ref{sechiex}),
and asymptotic results for higher-order observables on large systems
(section~\ref{sechiasy}).
The occupation probability falls off as $p_n\sim n^{-1/b}$,
whereas the rescaled variable
\beq
\nu=\frac{N}{L^{(b-1)/b}}
\eeq
has a non-trivial limit law $\P(\nu)$, depending on $b$.
The mean number of blockages grows as $\ln L$, with unit prefactor,
irrespective of $b$.
Section~\ref{disc} contains a brief outline of our findings.
An appendix is devoted to the linear recursions obeyed by the single-site and
pair occupation probabilities in the case where $b=2$.

\section{The model with $b=1$ and the statistics of records}
\label{secun}

The case where $b=1$ can be solved exactly by means of a mapping onto sequences of records.
The following solution serves as a warming-up exercise,
before we tackle the more intricate case of higher $b$,
where steric interactions induce non-trivial long-range correlations.

This model has already been investigated in~\cite{greek},
where the results~(\ref{averec}) and~(\ref{ave2s})
on the mean number of particles are derived.
Reference~\cite{greek} also deals with a generalization of the model
where only a fraction $p$ of the theatergoers are selfish and block further sites,
while the others are courteous and let latecomers pass them.

\begin{figure}[!ht]
\begin{center}
\includegraphics[angle=0,width=.6\linewidth]{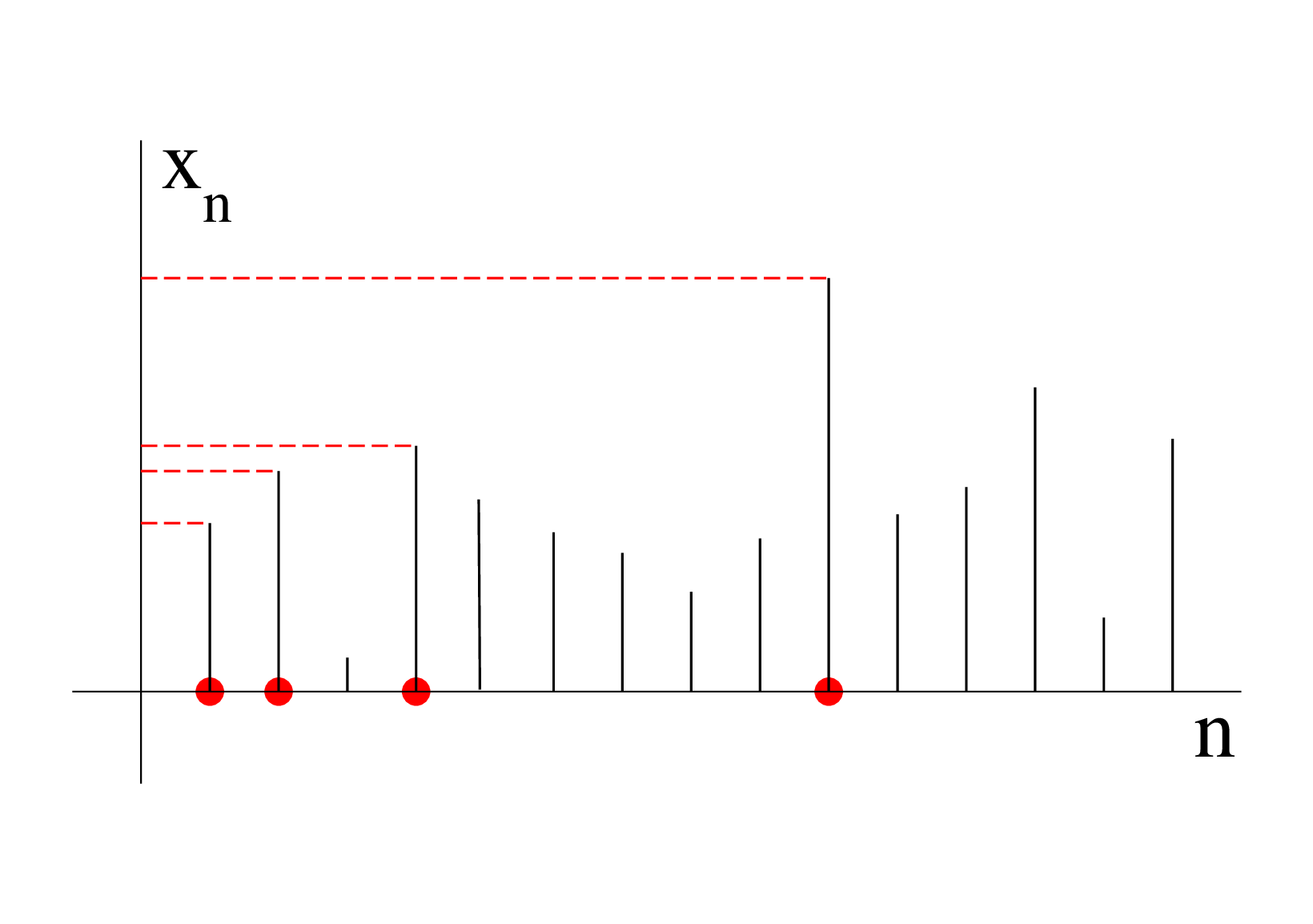}
\caption{\small
A randomly chosen height profile over a system of length $L=15$.
Symbols show the records of the height process.
Sites 10, 4, 2 and 1 become successively occupied during the process.}
\label{recs}
\end{center}
\end{figure}

Here, any occupied site blocks all sites to its right.
In other words, at any instant of time,
available sites are only those preceding the first occupied one.
The process stops when the first site is occupied.
If site $n$ remains empty, this means that at least one site to its left
has been visited and occupied before site $n$ was visited.
In terms of the height variables,
site $n$ is occupied in a jammed configuration
if and only if all the sites to its left have smaller heights,
i.e., $x_m<x_n$ for $m=1,\dots,n-1$.
The height $x_n$ at site $n$ therefore breaks the current record height.
Site $n$ is said to be a record~\cite{chan,ren} of the height process.
During the course of the process,
record sites become successively occupied in an ordered way from right to left.
This is illustrated in figure~\ref{recs},
showing a randomly chosen height profile and the corresponding records.

The occupied sites in a jammed configuration ---of any length $L$---
are therefore distributed as the records in a sequence of independent and
identically distributed random variables.
There is a vast literature on the statistics of records (see~\cite{gli,arn,nev} for reviews).
Here we provide a brief self-consistent account of the results which are
relevant to the present purpose.
The probability that site $n$ is occupied (i.e., a record) is exactly
\beq
p_n=\frac{1}{n}.
\label{precs}
\eeq
The largest among the first $n$ values $x_j$ is indeed any of them
with equal probabilities.
The most remarkable feature of the record process is that the presence of a record
at position $n$ is independent of the positions of all other records.
In other words, sites are independently occupied with probabilities $p_n$ given by~(\ref{precs}).

The mean number of particles (i.e., records) in a system of size $L$ therefore reads
\beq
\mean{N}=\sum_{n=1}^L\frac{1}{n}=H_L=\ln L+\gamma+\cdots,
\label{averec}
\eeq
where $H_L$ is the $L$th harmonic number,
$\gamma=0.577215\dots$ is Euler's constant,
and subleading terms go to zero.

As the occupations of different sites are statistically independent,
the generating function of the full distribution of $N$ reads
\beqa
G_L(z)&=&\mean{z^N}=\prod_{n=1}^L(zp_n+1-p_n)
=\prod_{n=1}^L\frac{z+n-1}{n}
\nonumber\\
&=&\frac{\Gamma(z+L)}{L!\,\Gamma(z)}
=\frac{1}{L!}\sum_{N=1}^L\stir{L}{N}z^N,
\label{zrec}
\eeqa
where $\stir{L}{N}$,
the Stirling number of the first kind~\cite{stir},
is the number of permutations of~$L$ objects having $N$ cycles.
There is indeed a correspondence
between records and cycles of permutations~\cite{ren}
(see also~\cite{feller,knuth,GKP,FS}).

The probability of having $N$ particles in a system of size $L$ therefore reads
\beq
P(N,L)=\frac{1}{L!}\stir{L}{N}\qquad(N=1,\dots,L).
\eeq
This distribution plays a role in various kinds of models.
It describes e.g.~the outcome of the ballistic aggregation process
where particles undergo totally inelastic collisions~\cite{IBA1,IBA2,IBA3,majumdar}.
It also arises in studies of leads and lead changes in growing networks~\cite{KR02,JML}.

In particular, there is a single particle with probability
\beq
P(1,L)=\frac{1}{L},
\label{p1lrec}
\eeq
corresponding to height profiles where $x_1$ is the largest,
i.e., to histories where the first site is visited and occupied first, blocking all other ones.
The other extreme situation where the system ends up entirely filled
occurs with the much smaller probability
\beq
P(L,L)=\frac{1}{L!},
\label{pllrec}
\eeq
corresponding to the ordering $x_1<x_2<\dots<x_L$,
so that each site is a record,
i.e., to histories such that $L$ is visited and occupied first, then $L-1$, and so on,
until the first site is visited last.

The distribution of the number of particles is self-averaging.
Setting $z=\e^s$ in the expression~(\ref{zrec}) of $G_L(z)$,
we indeed find that all cumulants of $N$
grow logarithmically with $L$, with unit prefactor, i.e.,
\beq
\mean{N^k}_c=\ln L+a_k+\cdots,
\label{cumrec}
\eeq
where subleading terms go to zero.
In other words, to leading order for large $L$, the distribution of $N$
becomes asymptotically a Poissonian distribution with parameter $\lambda=\ln L$.
The correction terms $a_k$ are numerical constants such that
\beq
\sum_{k\ge1}\frac{a_k}{k!}\,s^k=-\ln\Gamma(\e^s),
\eeq
i.e.,
\beqa
a_1=\gamma=0.577215\dots,
\nonumber\\
a_2=\gamma-\frac{\pi^2}{6}=-1.067718\dots,
\nonumber\\
a_3=\gamma-\frac{\pi^2}{2}+2\zeta(3)=-1.953472\dots,
\eeqa
and so on.

\subsection*{An apart\'e on the two-sided variant of the model}
\label{sec2s}

Before we investigate the more intricate cases of higher $b$,
it is worth considering the two-sided variant of the present model with $b=1$,
where particles may enter the system from either end,
whereas any occupied site blocks all further sites,
as viewed by the incoming particle.
This variant of the model was also considered in~\cite{greek}.
It can still be solved exactly,
although the occupations of different sites are not independent any more.

The solution of the two-sided model goes as follows.
The first particle enters from either end,
and occupies site $K$, chosen uniformly in the range $K=1,\dots,L$.
The system is thus divided into two subsystems of lengths $K-1$ and $L-K$, as shown here
\beq
\rightarrow\,\underbrace{\circ\,\circ\,\circ\,\circ\,\circ\,\circ\,\circ\,
\circ}_{K-1}\,\bullet\,
\underbrace{\circ\,\circ\,\circ\,\circ\,\circ\,\circ}_{L-K}\,\leftarrow
\eeq
for $L=15$ and $K=9$.
The subsequent history of each subsystem then follows the rules
of the one-sided model, studied above.
Assuming the system size is $L\ge2$,
the first and the last sites are both occupied
in jammed configurations of the two-sided model,
so that the particle number obeys $N\ge2$.

The occupation probabilities can be derived by conditioning on the
position $K$ of the first particle.
For the probability $p_n$ that site $n$ is occupied, this reads
\beq
p_n=\frac{1}{L}\sum_{K=1}^Lp_n(K),
\label{pnk}
\eeq
where the conditional probabilities $p_n(K)$ are as follows:
\beq
\left\{\matrix{
K=1,\dots,n-1:\hfill & p_n(K)=\frad{1}{L+1-n},\hfill\cr\cr
K=n:\hfill & p_n(K)=1,\hfill\cr\cr
K=n+1,\dots,L:\hfill & p_n(K)=\frad{1}{n},\hfill
}\right.
\eeq
and so
\beq
p_n=\frac{1}{n}+\frac{1}{L+1-n}-\frac{1}{L}.
\label{pn2s}
\eeq
At variance with~(\ref{precs}), this expression depends on both $L$ and $n$.
The first two terms can be viewed as the contributions of particles entering
from either end, whereas the last one is a non-trivial finite-size correction.
We have $p_1=p_L=1$, as should be.

The mean number of particles therefore reads
\beqa
\mean{N}=2H_L-1=2\ln L+2\gamma-1+\cdots,
\label{ave2s}
\eeqa
where subleading terms go to zero.

For the joint occupation probability $p_{m,n}$ (with $m<n$), we have similarly
\beq
\left\{\matrix{
K=1,\dots,m-1:\hfill & p_{m,n}(K)=\frad{1}{(L+1-m)(L+1-n)},\hfill\cr\cr
K=m:\hfill & p_{m,n}(K)=\frad{1}{L+1-n},\hfill\cr\cr
K=m+1,\dots,n-1:\!\!\!\!\!& p_{m,n}(K)=\frad{1}{m(L+1-n)},\hfill\cr\cr
K=n:\hfill & p_{m,n}(K)=\frad{1}{m},\hfill\cr\cr
K=n+1,\dots,L:\hfill & p_{m,n}(K)=\frad{1}{mn},\hfill
}\right.
\eeq
and so
\beqa
p_{m,n}
&=&\frac{1}{m}\left(\frac{1}{n}+\frac{1}{L+1-n}-\frac{1}{L}\right)
+\frac{1}{L+1-n}\left(\frac{1}{L+1-m}-\frac{1}{L}\right)
\nonumber\\
&=&p_mp_n-\frac{(m-1)(L-n)}{L^2(L+1-m)n}.
\label{pmn2s}
\eeqa
The second expression demonstrates that the occupations are negatively correlated,
whereas they were independent in the one-sided case.
The factors $m-1$ and $L-n$ reflect the property
that the first and last sites are always occupied.

The generating function of the full distribution of $N$ can also be evaluated as follows:
\beqa
G_L(z)&=&\mean{z^N}
\nonumber\\
&=&\frac{z}{L}\sum_{K=1}^L
\frac{\Gamma(z+K-1)}{(K-1)!\,\Gamma(z)}
\,\frac{\Gamma(z+L-K)}{(L-K)!\,\Gamma(z)}
\nonumber\\
&=&\frac{z}{L\,\Gamma(z)^2}\sum_{K=1}^L
\int_0^\infty\frac{x^{z+K-2}}{(K-1)!}\e^{-x}\dd x
\!\int_0^\infty\frac{y^{z+L-K-1}}{(L-K)!}\e^{-y}\dd y
\nonumber\\
&=&\frac{z}{L!\,\Gamma(z)^2}
\int_0^\infty x^{z-1}\e^{-x}\dd x
\int_0^\infty (x+y)^{L-1} y^{z-1}\e^{-y}\dd y
\nonumber\\
&=&\frac{z}{L!\,\Gamma(2z)}
\int_0^\infty s^{2z+L-2}\e^{-s}\dd s
\nonumber\\
&=&\frac{z\Gamma(2z+L-1)}{L!\,\Gamma(2z)},
\label{z2s}
\eeqa
where the second line is obtained by conditioning on the value of $K$,
as in~(\ref{pnk}), and using~(\ref{zrec}),
the third line is obtained by introducing integral expressions
for the $\Gamma$ functions in the numerators,
the fourth line is obtained by performing a binomial sum over $K-1=0,\dots,L-1$,
and the fifth line is the outcome of integrating over $x$ at fixed sum $s=x+y$.

The probability of having $N$ particles in a system of size $L$ therefore reads
\beq
P(N,L)=\frac{2^{N-1}}{L!}\stir{L-1}{N-1}\qquad(N=2,\dots,L),
\eeq
where $\stir{L-1}{N-1}$ is again the Stirling number of the first kind.
In particular, the system contains only two particles at its endpoints with probability
\beq
P(2,L)=\frac{2}{L(L-1)},
\label{p2lrec}
\eeq
whereas the other extreme situation where the system is entirely filled occurs
with probability
\beq
P(L,L)=\frac{2^{L-1}}{L!}.
\eeq
This probability is exponentially larger than in the one-sided case
(see~(\ref{pllrec})), but still factorially decaying.

The distribution of the number of particles is again self-averaging,
as all its cumulants grow logarithmically with $L$, i.e.,
\beq
\mean{N^k}_c=2\ln L+b_k+\cdots
\label{cum2s}
\eeq
In other words, to leading order for large $L$, the distribution of $N$
becomes asymptotically a Poissonian distribution with parameter $\lambda=2\ln L$.
The correction terms $b_k$ are numerical constants such that
\beq
\sum_{k\ge1}\frac{b_k}{k!}\,s^k=s-\ln\Gamma(2\e^s),
\eeq
i.e.,
\beqa
b_1=2\gamma-1=0.154431\dots,
\nonumber\\
b_2=2\gamma+2-\frac{2\pi^2}{3}=-3.425304\dots,
\nonumber\\
b_3=2\gamma-6-2\pi^2+16\zeta(3)=-5.351867\dots,
\eeqa
and so on.

\section{The model with $b=2$}
\label{secde}

In this section we investigate the theater model {\it stricto sensu} ($b=2$) in full detail.
In terms of the height variables,
site $n$ is occupied in a jammed configuration
if and only if there is no pair of consecutive sites before it with larger heights,
i.e., no integer $m=1,\dots,n-2$ such that $x_m>x_n$ and $x_{m+1}>x_n$.
This condition only depends on the ordering of the $L$ height variables,
i.e., equivalently, on the ordering of the times of first visits to the $L$ sites of the system.

Let us begin with an explicit solution of the problem for a system of size $L=4$
by enumerating all cases.
Table~\ref{four} gives a list of the $4!=24$ equally probable orderings
of the four height variables.
Underlined figures stand for occupied sites in the corresponding jammed configuration.
Site 3 is occupied in 16 cases and site 4 is occupied in 14 cases,
whereas both of them are occupied in 10 cases and none of them in 4 cases.

\begin{table}[!ht]
\begin{center}
\begin{tabular}{|c|c|c|c|}
\hline
\u1 \u2 3 4 & \u2 \u1 3 4 & \u3 \u1 \u2 4 & \u4 \u1 \u2 3 \\
\u1 \u2 4 3 & \u2 \u1 4 3 & \u3 \u1 \u4 \u2 & \u4 \u1 \u3 \u2 \\
\u1 \u3 \u2 4 & \u2 \u3 \u1 4 & \u3 \u2 \u1 4 & \u4 \u2 \u1 3 \\
\u1 \u3 \u4 \u2 & \u2 \u3 4 \u1 & \u3 \u2 4 \u1 & \u4 \u2 \u3 \u1 \\
\u1 \u4 \u2 3 & \u2 \u4 \u1 3 & \u3 \u4 \u1 \u2 & \u4 \u3 \u1 \u2 \\
\u1 \u4 \u3 \u2 & \u2 \u4 \u3 \u1 & \u3 \u4 \u2 \u1 & \u4 \u3 \u2 \u1 \\
\hline
\end{tabular}
\caption{The $4!=24$ possible ordering of height variables for a system of size $L=4$.
Underlined figures stand for occupied sites in the corresponding jammed configuration.}
\label{four}
\end{center}
\end{table}

The full probability distribution of $N$ for $L=4$ therefore reads
\beq
P(2,4)=\frac{1}{6},\qquad
P(3,4)=P(4,4)=\frac{5}{12},
\eeq
whereas the occupation probabilities
\beq
p_3=\frac{2}{3},\qquad p_4=\frac{7}{12},\qquad p_{3,4}=\frac{5}{12}
\eeq
hold for all $L\ge4$.
The joint probability $p_{3,4}=15/36$ is larger than the product $p_3p_4=14/36$.
This demonstrates that occupations of different sites are positively correlated,
whereas they were independent in the case where $b=1$,
and negatively correlated in the two-sided variant of the latter model.

We shall successively derive exact results for finite systems (section~\ref{secdeex})
and asymptotic ones for large systems (section~\ref{secdeasy}).

\subsection{Exact results for finite systems}
\label{secdeex}

Let us first focus our attention onto occupation probabilities.
The occupation probability $p_n$ is the probability that there is no integer
in the range $m=1,\dots,n-2$ such that $x_m>x_n$ and $x_{m+1}>x_n$.
This quantity can be derived as follows.
Set $x_n=1-y$, and consider an auxiliary problem
where each site of the system is independently occupied with probability $y$.
Let $f_m(y)$ be the probability that no pair of consecutive sites
is occupied among the first $m$ sites.
We have then
\beq
p_n=\int_0^1 f_{n-1}(y)\,\dd y.
\label{pnint}
\eeq
In order to proceed, we write $f_m(y)=f_m^\bullet(y)+f_m^\circ(y)$,
where $f_m^\bullet(y)$ (resp.~$f_m^\circ(y)$)
corresponds to allowed configurations where site $m$ is occupied (resp.~empty).
These quantities obey the recursions
\beq
\left\{\matrix{
f_m^\bullet(y)=yf_{m-1}^\circ(y),\hfill\cr\cr
f_m^\circ(y)=(1-y)\left(f_{m-1}^\bullet(y)+f_{m-1}^\circ(y)\right),\cr
}\right.
\eeq
and so
\beq
f_m(y)=(1-y)f_{m-1}(y)+y(1-y)f_{m-2}(y),
\label{fmrec}
\eeq
with $f_0(y)=f_1(y)=1$.
The $f_m(y)$ are polynomials in $y$ with increasing degrees.
Looking for a solution to~(\ref{fmrec}) of the form
\beq
f_m(y)=\sum_ka_{m,k}y^k(1-y)^{m-k},
\label{fmres}
\eeq
we find that the coefficients $a_{m,k}$ obey the recursion
\beq
a_{m,k}=a_{m-1,k}+a_{m-2,k-1},
\eeq
with initial condition $a_{0,0}=1$,
whose solution reads
\beq
a_{m,k}=\frac{(m+1-k)!}{k!(m+1-2k)!}\qquad(k=0,\dots,\Int{(m+1)/2}).
\label{amkres}
\eeq
Inserting~(\ref{fmres}) and~(\ref{amkres}) into~(\ref{pnint})
and working out the integral, we obtain
\beq
p_n=\frac{A_n}{n!},
\label{pres}
\eeq
with
\beq
A_n=\sum_{k=0}^{\Int{n/2}}\frac{(n-k)!(n-k-1)!}{(n-2k)!}.
\eeq
The mean particle number then reads (see~(\ref{ave}))
\beq
\mean{N}=\frac{B_L}{L!},
\label{averes}
\eeq
with
\beq
B_L=\sum_{n=1}^L\frac{L!}{n!}\,A_n.
\eeq
The integers $A_n$ and $B_n$ are listed up to $n=10$ in table~\ref{ints}.

\begin{table}[!ht]
\begin{center}
\begin{tabular}{|r|r|r|r|r|}
\hline
$n$ & $A_n$ & $B_n$ & $C_n$ & $D_n$\\
\hline
1 & 1 & 1 & 2 & 1\\
2 & 2 & 4 & 4 & 2\\
3 & 4 & 16 & 10 & 4\\
4 & 14 & 78 & 38 & 10\\
5 & 60 & 450 & 180 & 26\\
6 & 324 & 3\,024 & 1\,044 & 76\\
7 & 2\,064 & 23\,232 & 7\,104 & 232\\
8 & 15\,264 & 201\,120 & 55\,584 & 764\\
9 & 128\,160 & 1\,938\,240 & 491\,040 & 2\,620\\
10 & 1\,205\,280 & 20\,587\,680 & 4\,834\,080 & 9\,496\\
\hline
\end{tabular}
\caption{The integers $A_n$, $B_n$, $C_n$ and $D_n$ up to $n=10$.
These quantities respectively enter the exact expressions
(\ref{pres}) for the occupation probability~$p_n$,
(\ref{averes}) for the mean particle number $\mean{N}$,
(\ref{ppres}) for the pair occupation probability $p_{n,n+1}$
and~(\ref{pllres}) for the probability $P(L,L)$ of densest configurations.}
\label{ints}
\end{center}
\end{table}

In order to investigate asymptotic properties of the above quantities
at large $n$ or $L$, it is advantageous to use generating functions.
The generating function of the $f_m(y)$
can be derived from the recursion~(\ref{fmrec}).
It reads
\beq
F(y,z)=\sum_{m\ge0}f_m(y)z^m=\frac{1+yz}{1+(y-1)z+y(y-1)z^2}.
\label{fyzres}
\eeq
The generating function of the occupation probabilities reads (see~(\ref{pnint}))
\beqa
\Pi(z)
&=&\sum_{n\ge1}p_nz^n
\nonumber\\
&=&z\int_0^1F(y,z)\,\dd y
\nonumber\\
&=&\frac{1+z}{\sqrt{(1-z)(3+z)}}
\arctan\left(\frac{z}{2+z}\sqrt{\frac{3+z}{1-z}}\right)
-\frac{\ln(1-z)}{2}.
\label{pires}
\eeqa
The behavior of the latter expression near $z=1$ governs the behavior of $p_n$ at large~$n$.
Setting $z=\e^{-\eps}$, the expansion
\beq
\Pi(z)=\frac{\pi}{2\sqrt{\eps}}-\frac{1}{2}(\ln\eps+3)
-\frac{\pi\sqrt\eps}{16}+\cdots
\eeq
translates to
\beq
p_n=\frac{\sqrt{\pi}}{2\sqrt{n}}+\frac{1}{2n}
+\frac{\sqrt{\pi}}{32n\sqrt{n}}+\cdots
\label{pnasy}
\eeq
and
\beq
\mean{N}=\sqrt{\pi L}+\frac{1}{2}(\ln
L+\gamma-3)+\frac{3\sqrt{\pi}}{16\sqrt{L}}+\cdots
\label{aveasy}
\eeq
To leading order, the decay of the occupation probability $p_n$
and the growth of the mean particle number $\mean{N}$ are described by simple power laws.
The occupation probabilities $p_n$ will be shown in~\ref{app} to obey the recursion~(\ref{prec}),
allowing one to systematically derive more terms of the
expansions~(\ref{pnasy}) and~(\ref{aveasy}).

The above technique can be extended to higher-order occupation probabilities
such as $p_{m,n}$.
The resulting expressions however soon become very cumbersome.
We shall focus our attention onto the pair occupation probability $p_{n,n+1}$,
i.e., the probability that two successive sites end up being simultaneously occupied.
This quantity can be expressed in terms of the sole probabilities $f_m(y)$ introduced above.
We obtain after some algebra
\beq
p_{n,n+1}=\int_0^1\bigl(f_{n-1}(y)+(1-y)f_{n-2}(y)\bigr)y\,\dd y,
\label{ppint}
\eeq
where the first (resp.~second) term inside the large parentheses corresponds to histories
where site $n+1$ is visited and occupied before (resp.~after) site $n$.
Inserting~(\ref{fmres}) and~(\ref{amkres}) into~(\ref{ppint})
and working out the integral, we obtain
\beq
p_{n,n+1}=\frac{C_n}{(n+1)!},
\label{ppres}
\eeq
with
\beq
C_n=\sum_{k=0}^{\Int{n/2}}\frac{(k+1)(2n-3k)(n-k-1)!^2}{(n-2k)!}.
\eeq
The integers $C_n$ are listed up to $n=10$ in table~\ref{ints}.

The generating function of the pair occupation probabilities reads
\beqa
\Pi_1(z)
&=&\sum_{n\ge1}p_{n,n+1}z^n
\nonumber\\
&=&z\int_0^1(F(y,z)+1+(1-y)zF(y,z))y\,\dd y
\nonumber\\
&=&z\int_0^1\frac{2+yz}{1+(y-1)z+y(y-1)z^2}\,y\,\dd y
\nonumber\\
&=&1-\frac{1+z}{2z}\ln(1-z)
\nonumber\\
&-&\frac{\sqrt{(1-z)(3+z)}}{z}
\arctan\left(\frac{z}{2+z}\sqrt{\frac{3+z}{1-z}}\right).
\label{pi1res}
\eeqa
Setting again $z=\e^{-\eps}$, the expansion
\beq
\Pi_1(z)=1-\ln\eps-\pi\sqrt{\eps}+\frac{\eps}{2}(7-\ln\eps)+\cdots
\eeq
translates to
\beq
p_{n,n+1}=\frac{1}{n}+\frac{\sqrt{\pi}}{2n\sqrt{n}}-\frac{1}{2n^2}+\cdots
\label{ppasy}
\eeq
and
\beq
\sum_{n=1}^{L-1}p_{n,n+1}=\ln L+\gamma+1-\frac{\sqrt{\pi}}{\sqrt{L}}+\cdots
\label{pairasy}
\eeq
The latter quantity is nothing but the mean number of blockages on a system of size~$L$.

To leading order, we have $p_{n,n+1}\approx1/n$,
whereas the corresponding product of single occupation probabilities
reads $p_np_{n+1}\approx\pi/(4n)$ (see~(\ref{pnasy})).
In other words, the correlation between occupations of pairs of neighboring sites
results in the asymptotic enhancement factor
\beq
\frac{p_{n,n+1}}{p_np_{n+1}}\to\frac{4}{\pi}=1.273239\dots
\label{q2asy}
\eeq
The pair occupation probabilities $p_{n,n+1}$ will be shown in~\ref{app} to
obey the recursion~(\ref{pprec}),
allowing one to systematically derive more terms of the expansion~(\ref{ppasy}).

Our next goal is to derive exact expressions for the probability
that the system ends up either in the least dense or the densest configurations.
Configurations where only the first two sites are occupied
(see~(\ref{min:jammed}))
correspond to height profiles where $x_1$ and $x_2$ are the two largest values.
Their probability reads
\beq
P(2,L)=\frac{2}{L(L-1)}.
\label{p2lres}
\eeq
The probability of the other extreme situation where the system is entirely filled
(see~(\ref{max:jammed})) can also be worked out exactly.
Setting
\beq
P(L,L)=\frac{D_L}{L!},
\label{pllres}
\eeq
the numbers $D_L$ of permutations such that the system ends up entirely filled
can be determined recursively as follows.
The site which is occupied last must be either the first or the second one.
The number of permutations such that the first site is occupied last is $D_{L-1}$.
The number of permutations such that the second site is occupied last is $(L-1)D_{L-2}$,
where the factor $L-1$ counts the number of ways of inserting site 1 in a
permutation of the $L-2$ sites $n=3,\dots,L$.
Hence the recursion
\beq
D_L=D_{L-1}+(L-1)D_{L-2},
\label{drec}
\eeq
with $D_0=D_1=1$.
The exponential generating function
\beq
\Delta(z)=\sum_{L\ge0}\frac{D_L}{L!}z^L
\eeq
obeys the differential equation $\Delta'(z)=(z+1)\Delta(z)$, hence
\beq
\Delta(z)=\exp\left(z+\frac{z^2}{2}\right),
\label{deltaz}
\eeq
and therefore
\beq
D_L=\sum_{k=0}^\Int{L/2}\frac{L!}{2^kk!(L-2k)!}.
\eeq
The integers $D_L$ are given in the OEIS~\cite{OEIS} as sequence A000085.
They have several combinatorial interpretations.
In particular, $D_L$ is the number of involutive permutations of $L$ objects,
i.e., permutations consisting of cycles of length at most 2.
The $D_L$ are listed up to $L=10$ in table~\ref{ints}.
Their asymptotic behavior can be derived from~(\ref{deltaz}):
\beq
\ln D_L=\frac{L}{2}(\ln L-1)+\sqrt{L}-\frac{\ln 2}{2}-\frac{1}{4}+\cdots
\eeq
This translates to
\beq
\ln P(L,L)=-\frac{L}{2}(\ln L-1)+\sqrt{L}-\frac{\ln 4\pi L}{2}
-\frac{1}{4}+\cdots
\label{pllas}
\eeq
The probability of densest configurations therefore exhibits a stretched factorial falloff.
This result can be put in perspective with the following heuristic picture.
An efficient way of building a completely filled configuration on an even-sized system
consists in filling first
all odd sites in whichever order ---there are $(L/2)!$ ways of doing so---
and then all even sites in an ordered way from right to left.
The resulting estimate, $P(L,L)\sim(L/2)!/L!$,
shares the same stretched factorial falloff as~(\ref{pllas}).

\subsection{Asymptotic results for large systems}
\label{secdeasy}

The formalism used in section~\ref{secdeex} to derive exact results on finite systems
simplifies for large systems,
to the extent that it becomes possible to evaluate the scaling behavior of more
intricate quantities,
such as higher-order occupation probabilities
and higher moments of the total number of particles.
The key point is the following.
The probabilities $f_m(y)$ introduced in the beginning of section~\ref{secdeex}
assume a simple exponential scaling form,\footnote{Throughout this paper,
the symbol $\approx$ denotes an asymptotic equality.}
\beq
f_m(y)\approx\e^{-my^2},
\label{fas}
\eeq
in the relevant regime where $m$ is large and $y$ is small.
From a technical standpoint, the above expression can be derived
by setting $z=\e^{-\eps}$ in~(\ref{fyzres}).
If $\eps$ and $y$ are simultaneously small, the latter expression becomes
\beq
F(y,z)\approx\frac{1}{\eps+y^2},
\eeq
which translates to~(\ref{fas}).
The result~(\ref{fas}) can be alternatively derived by means of heuristic reasoning.
If the occupation probability $y$ defining the auxiliary problem is small,
the density of pairs of consecutive particles is $y^2$, to leading order,
and so the mean number of such pairs among the first $m$ sites is approximately
$\lambda=my^2$.
Furthermore, the number of such pairs is expected to follow a Poissonian statistics
in this dilute regime.
The probability of having no pair is therefore approximately $\exp(-\lambda)$,
which is precisely~(\ref{fas}).

Within this framework, the occupation probability $p_n$
admits the following simple expression at large $n$:
\beq
p_n\approx\int_0^\infty\e^{-ny^2}\dd y=\frac{\sqrt{\pi}}{2\sqrt{n}}.
\eeq
The mean number of particles in a system of size $L$ therefore reads
\beq
\mean{N}\approx\mu_1\sqrt{L},
\eeq
with
\beq
\mu_1=\sqrt{\pi},
\label{mu1}
\eeq
in agreement with the leading-order behavior of the exact results~(\ref{pnasy})
and~(\ref{aveasy}).

The same setting applies to higher-order quantities as well.
The joint occupation probability $p_{m,n}$ thus reads
\beq
p_{m,n}\approx\int_0^\infty\int_0^\infty\e^{-mY^2-(n-m)y_2^2}\dd y_1\dd y_2,
\eeq
whenever $m$ and $n-m$ are simultaneously large.
The integration variables $y_1$ and $y_2$ respectively stand for $1-x_m$ and $1-x_n$, and
\beq
Y=\max(y_1,y_2)
\label{Ydef}
\eeq
is the larger of both of them.
The exponential factor expresses the constraints
that there is no pair of consecutive sites before $m$ whose height is higher than $x_m$
and no pair of consecutive sites before $n$ whose height is higher than $x_n$.
Adding up the contributions of the sectors where $y_1<y_2$ and $y_1>y_2$, we obtain
\beq
p_{m,n}\approx\frac{1}{2n}+\frac{1}{2\sqrt{m(n-m)}}\arctan\sqrt\frac{n-m}{m}.
\label{pmnasy}
\eeq

If both sites are very far apart, the expression~(\ref{pmnasy}) becomes
\beq
p_{m,n}\approx\frac{\pi}{4\sqrt{mn}}\approx p_mp_n\qquad(1\ll m\ll n),
\eeq
meaning that the occupations of sites $m$ and $n$ are asymptotically
uncorrelated in that regime.
In the opposite case where $m$ and $n$ are close to each other, we obtain
\beq
p_{m,n}\approx\frac{1}{n}\qquad(1\ll n-m\ll n).
\eeq
This expression matches the leading-order behavior of the exact
result~(\ref{ppasy}) for $m=n-1$.

The expression~(\ref{pmnasy})
for the joint occupation probability assumes a scaling law of the form
\beq
p_{m,n}\approx\frac{\phi_2(u)}{n}\qquad(0<u=m/n<1),
\label{phi2de}
\eeq
with
\beq
\phi_2(u)=\frac{1}{2}+\frac{1}{2\sqrt{u(1-u)}}\arctan\sqrt\frac{1-u}{u}.
\eeq
The second moment of the number of particles in a system of size $L$
therefore reads
\beq
\mean{N^2}\approx\mu_2\,L,
\eeq
with
\beq
\mu_2=2\int_0^1\phi_2(u)\,\dd u=1+\frac{\pi^2}{4}=3.467401\dots
\label{mu2}
\eeq
The corresponding variance reads
\beq
\mean{N^2}_c\approx c_2\,L,
\eeq
with
\beq
c_2=\mu_2-\mu_1^2=\frac{(\pi-2)^2}{4}=0.325808\dots
\eeq
Finally, the relative number variance
\beq
V=\frac{\mean{N^2}_c}{\mean{N}^2}
\label{vdef}
\eeq
has a non-trivial value
\beq
V=\frac{c_2}{\mu_1^2}=\frac{\mu_2}{\mu_1^2}-1=\frac{(\pi-2)^2}{4\pi}=0.103708\dots
\label{v2}
\eeq
in the limit of a very large system.
This implies in particular that the number of particles in a jammed configuration
keeps fluctuating and does not become self-averaging in the thermodynamic limit.

The triple occupation probability $p_{l,m,n}$ reads similarly
\beq
p_{l,m,n}\approx\int_0^\infty\int_0^\infty\int_0^\infty
\e^{-lY_1^2-(m-l)Y_2^2-(n-m)y_3^2}\dd y_1\dd y_2\dd y_3,
\eeq
whenever $l$, $m-l$ and $n-m$ are simultaneously large,
with
\beq
Y_1=\max(y_1,y_2,y_3),\qquad Y_2=\max(y_2,y_3).
\eeq
The above expression can be shown to assume a scaling law of the form
\beq
p_{l,m,n}\approx\frac{\phi_3(u,v)}{n^{3/2}}\qquad(0<u=l/n<v=m/n<1),
\label{phi3de}
\eeq
with
\beqa
\phi_3(u,v)
&=&\frac{\sqrt{\pi}}{4}
\Biggl(1+\frac{1}{v}+\frac{1}{\sqrt{u}(1+\sqrt{u})}
\nonumber\\
&+&\frac{1}{\sqrt{u(v-u)(1-v)}}
\arctan\frac{\sqrt{(v-u)(1-v)}}{v+\sqrt{u}}\Biggr).
\eeqa
The third moment of the number of particles reads
\beq
\mean{N^3}\approx\mu_3\,L^{3/2},
\eeq
with
\beq
\mu_3=4\int_0^1\dd u\int_u^1\dd
v\,\phi_3(u,v)=\frac{\sqrt{\pi}(\pi^2+15)}{6}=7.346704\dots
\label{mu3}
\eeq

The general structure of higher-order quantities emerges clearly from the above.
In particular, the $k$th moment of the total number of particles scales as
\beq
\mean{N^k}\approx\mu_k\,L^{k/2},
\eeq
where the first three prefactors $\mu_k$
have been derived in~(\ref{mu1}), (\ref{mu2}) and~(\ref{mu3}).
As a consequence, the full probability distribution $P(N,L)$ is expected to scale as
\beq
P(N,L)\approx\frac{\P(\nu)}{\sqrt{L}},
\label{prosca}
\eeq
whenever $N$ and $L$ are both large, with a fixed value of the combination
\beq
\nu=\frac{N}{\sqrt{L}}.
\eeq
The prefactors $\mu_k$ are nothing but the moments
of the non-trivial limit law $\P(\nu)$:
\beq
\int_0^\infty\nu^k\P(\nu)\,\dd\nu=\mu_k.
\eeq
Matching the scaling law~(\ref{prosca}) with the exact result~(\ref{p2lres}) for the
probability of the least dense configurations suggests that $\P(\nu)$ vanishes as
\beq
\P(\nu)\sim\nu^3\qquad(\nu\ll1),
\label{prosmall}
\eeq
whereas matching it with the asymptotic result~(\ref{pllas}) for the probability
of densest configurations suggests the behavior
\beq
\ln\P(\nu)\sim-\nu^2\ln\nu\qquad(\nu\gg1).
\label{prolarge}
\eeq

\begin{figure}[!ht]
\begin{center}
\includegraphics[angle=0,width=.48\linewidth]{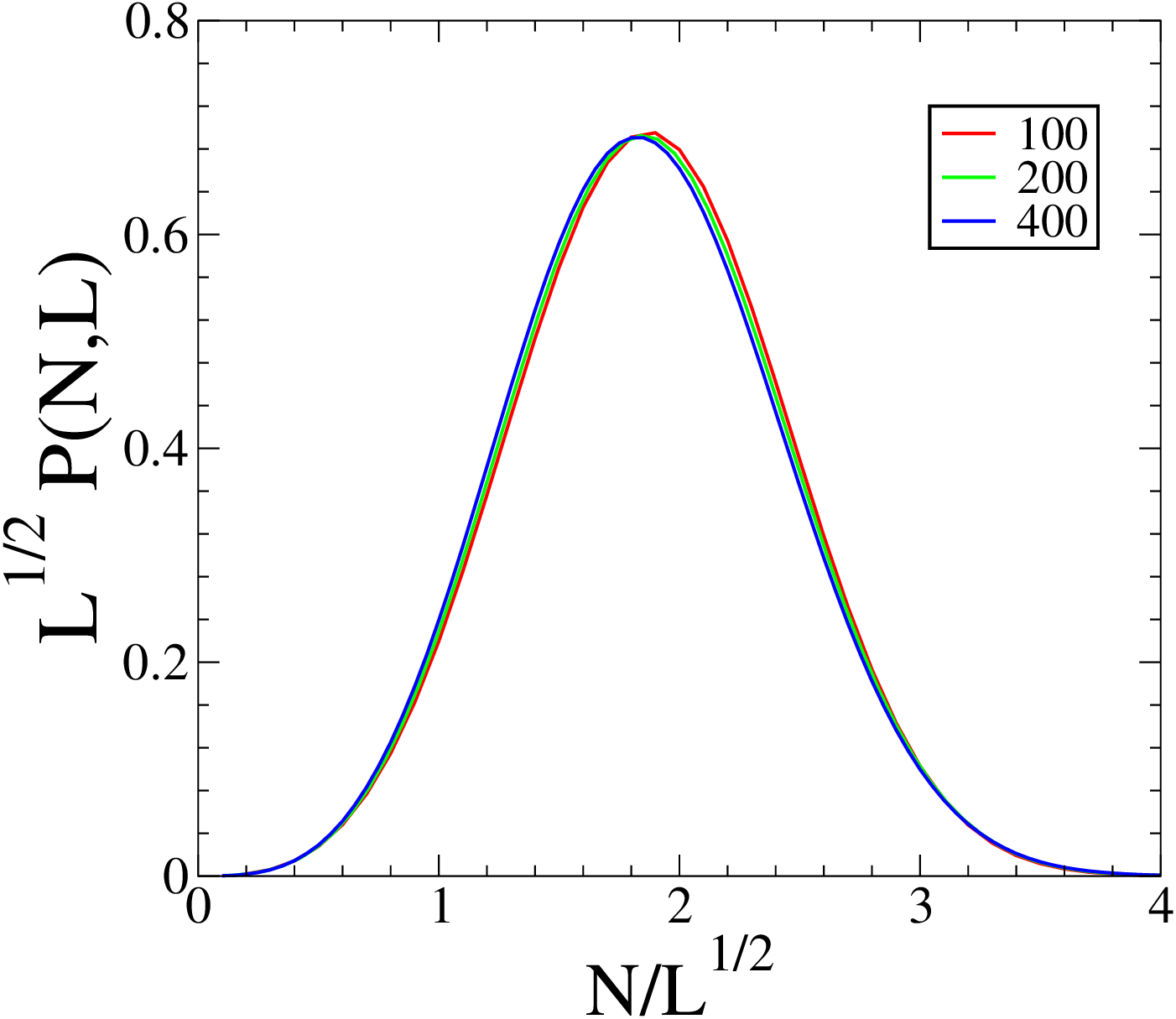}
\hskip 6pt
\includegraphics[angle=0,width=.48\linewidth]{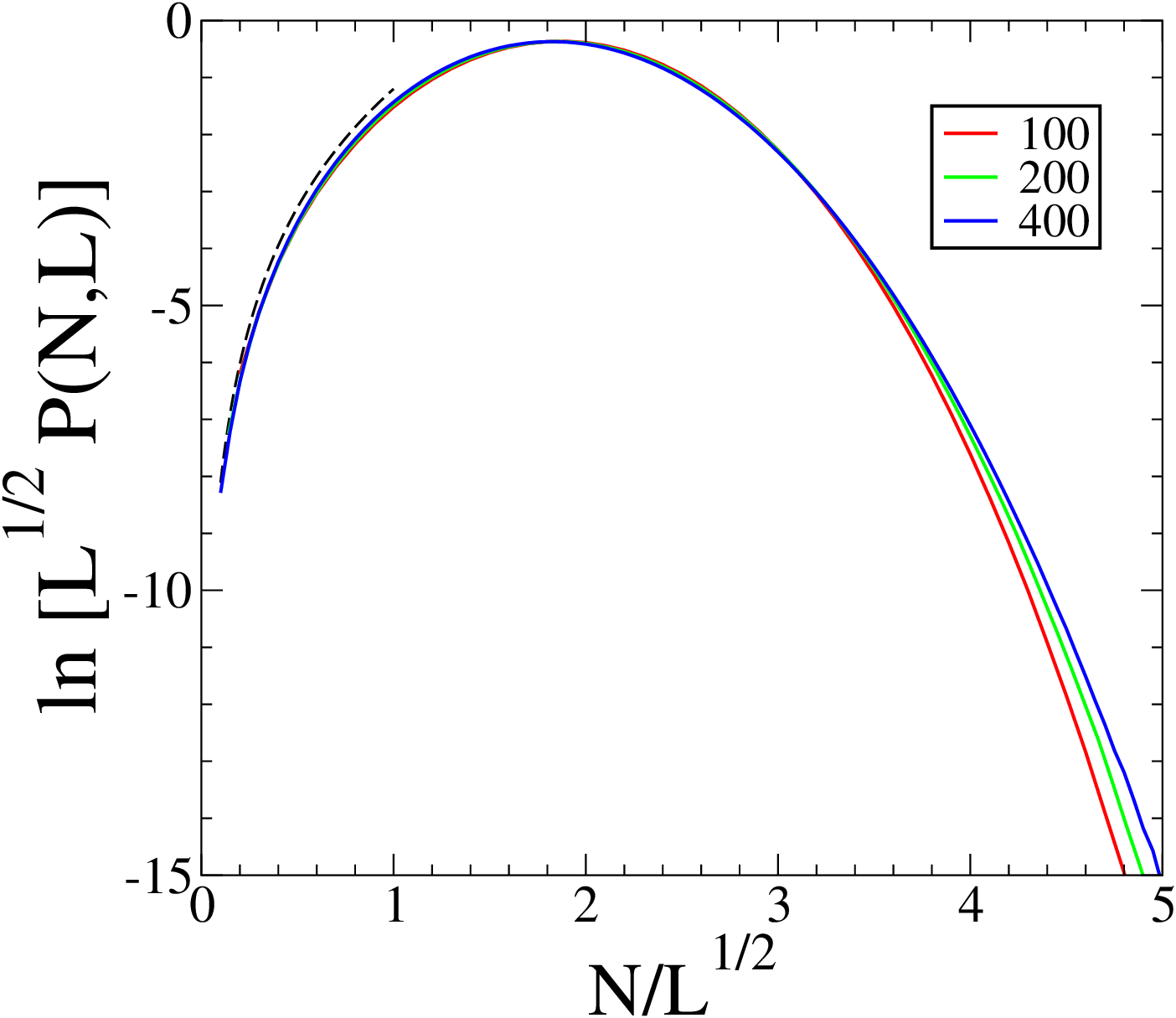}
\caption{\small
Plots of $\sqrt{L}\,P(N,L)$ in linear scale (left) and in logarithmic scale (right)
against $N/\sqrt{L}$, for system sizes $L=100$, 200 and 400.
The dashed line in the left part of the right panel shows the cubic law
predicted in~(\ref{prosmall}).}
\label{proplot}
\end{center}
\end{figure}

The scaling law~(\ref{prosca}) is illustrated in figure~\ref{proplot},
showing plots of $\sqrt{L}\,P(N,L)$ in linear scale (left) and in logarithmic
scale (right) against $N/\sqrt{L}$, for three different system sizes $L$.
Each dataset is obtained by means of a direct simulation
of $10^{10}$ histories of the process.
The existence of a limit law $\P(\nu)$ is corroborated by the good collapse
observed on the left panel.
The right panel emphasizes both the validity of the power-law
behavior~(\ref{prosmall})
at small $\nu$ ---the dashed line shows an exact $\nu^3$ law---
as well as the presence of more appreciable finite-size corrections at large~$\nu$.

To close, we consider the statistics of clusters of consecutive occupied sites
in jammed configurations.
Let $M_k$ be the number of clusters consisting of exactly $k$ sites on a system of size $L$.
The mean values of these numbers obey the sum rules
\beqa
\sum_{k\ge1}k\mean{M_k}=\mean{N},
\nonumber\\
\sum_{k\ge1}(k-1)\mean{M_k}=\sum_{n=1}^{L-1}p_{n,n+1},
\label{sumrules}
\eeqa
and so on.
It is clear from the above that only $\mean{M_1}$ and $\mean{M_2}$ diverge with the system size,
whereas the $\mean{M_k}$ converge to finite limits for all $k\ge3$.
Inserting the expansions~(\ref{aveasy}) and~(\ref{pairasy})
into the above sum rules yields
\beqa
\mean{M_1}=\sqrt{\pi L}-\frac{3}{2}\ln L+\cdots,
\nonumber\\
\mean{M_2}=\ln L+\cdots,
\eeqa
where the dots stand for numerical constants
which cannot be predicted by the above line of reasoning.

\section{The model with higher $b$}
\label{sechi}

This section is devoted to the extended theater model
where $b$ consecutive occupied sites are needed to constitute a blockage,
with $b$ being any integer in the range $b\ge2$.
The first $b$ sites are occupied in all jammed configurations.
In terms of the height variables, site $n\ge b+1$ is occupied if and only if
there is no integer $m=1,\dots,n-b$ such that $x_{m+k}>x_n$ for $k=0,\dots,b-1$.
Here again, this condition only depends on the ordering of the $L$ height variables.

Hereafter we assume that the system size $L$ is at least $b$.
We extend to higher values of $b$ some of the main outcomes
of the detailed investigation of the case where $b=2$ performed in section~\ref{secde}.
We successively derive exact results for finite systems (section~\ref{sechiex})
and asymptotic ones for large systems (section~\ref{sechiasy}).

\subsection{Exact results for finite systems}
\label{sechiex}

Exact expressions for the probability that the system ends up either in the least dense ($N=b$)
or the densest ($N=L$) configurations can be derived for all $b$.

The least dense configurations,
where only the first $b$ sites are occupied,
are in correspondence with height profiles where the $b$ largest values
are reached on the first~$b$ sites ($n=1,\dots,b$).
Their probability reads
\beq
P(b,L)=\frac{b!(L-b)!}{L!}.
\label{pblres}
\eeq
This result generalizes~(\ref{p1lrec}) and~(\ref{p2lres}).

The probability of densest configurations can also be worked out exactly,
along the lines of section~\ref{secdeex}.
Setting again (see~(\ref{pllres}))
\beq
P(L,L)=\frac{D\b_L}{L!},
\label{pbllres}
\eeq
the numbers $D\b_L$ can be shown to obey the recursion
\beq
D\b_L=\sum_{k=1}^b\frac{(L-1)!}{(L-k)!}\,D\b_{L-k},
\label{dbrec}
\eeq
with the (formal) initial values $D\b_L=L!$ for $L=1,\dots,b$.
The exponential generating function of the $D\b_L$ is found to be
\beq
\Delta\b(z)=\sum_{L\ge0}\frac{D\b_L}{L!}\,z^L=\exp\sum_{k=1}^b\frac{z^k}{k}.
\label{bdeltaz}
\eeq
The integers $D\b_L$ are given in the OEIS~\cite{OEIS}
as sequences A000085, A057693, A070945, A070946 and A070947 for $b=2$ to 6.
Quite generally, $D\b_L$ is the number of permu\-tations of $L$ objects
consisting of cycles of length at most $b$
(see~\cite{FS}, and~\cite{dissert} and the references therein).
The asymptotic behavior of the $D\b_L$ to leading order can be derived
from~(\ref{bdeltaz}):
\beq
\ln D\b_L\approx\frac{b-1}{b}\,L(\ln L-1),
\eeq
This translates to
\beq
\ln P(L,L)\approx-\frac{L}{b}(\ln L-1).
\label{pbllas}
\eeq

\subsection{Asymptotic results for large systems}
\label{sechiasy}

The approach of section~\ref{secdeasy} can be extended to higher values of $b$,
allowing one to evaluate the scaling behavior
of joint occupation probabilities and of moments of the total number of particles.
The starting point again consists
in considering the auxiliary problem where each site of the semi-infinite chain
is occupied with given probability~$y$.
Let $f_m(y)$ be the probability that there is no sequence of $b$ consecutive
occupied sites among the first $m$ sites.
In the relevant regime where $m$ is large and $y$ is small,
the above probability obeys an exponential scaling law:
\beq
f_m(y)\approx\e^{-my^b}.
\label{bfas}
\eeq
This expression, which generalizes~(\ref{fas}) to higher $b$,
can be derived by means of a similar heuristic reasoning based on Poissonian statistics.

The occupation probability $p_n$ has the following asymptotic expression at large $n$
(see~(\ref{pnint})):
\beq
p_n\approx\int_0^\infty\e^{-ny^b}\dd y=\frac{\Gamma(\frac{1}{b}+1)}{n^{1/b}}.
\eeq
This slow power-law falloff implies that the mean number of particles
grows as a subextensive power law of the system size $L$ for all integers $b\ge2$, i.e.,
\beq
\mean{N}\approx\mu_1L^{(b-1)/b},
\eeq
with
\beq
\mu_1=\frac{\Gamma(\frac{1}{b})}{b-1}.
\label{bmu1}
\eeq

Similarly, the joint occupation probability $p_{m,n}$ reads asymptotically
\beq
p_{m,n}\approx\int_0^\infty\int_0^\infty\e^{-mY^b-(n-m)y_2^b}\dd y_1\dd y_2,
\eeq
whenever $m$ and $n-m$ are simultaneously large,
with $Y=\max(y_1,y_2)$ (see~(\ref{Ydef})).
Some algebra allows one to recast this expression as a scaling law of the form
\beq
p_{m,n}\approx\frac{\phi_2(u)}{n^{2/b}}\qquad(0<u=m/n<1),
\eeq
with
\beq
\phi_2(u)=\frac{\Gamma(\frac{2}{b})}{b}
\left(1+\int_0^1\left(u+(1-u)\xi^b\right)^{-2/b}\dd\xi\right).
\eeq
The second moment of the number of particles therefore grows as
\beq
\mean{N^2}\approx\mu_2\,L^{2(b-1)/b},
\eeq
with
\beq
\mu_2=\frac{b}{b-1}\int_0^1\phi_2(u)\,\dd u
=\frac{\Gamma(\frac{2}{b})}{b-1}
\left(1+\frac{\pi}{b-2}\cot\frac{\pi}{b}\right).
\label{bmu2}
\eeq
Hence the relative number variance (see~(\ref{vdef})) has a non-trivial limiting value
\beq
V=\frac{\mu_2}{\mu_1^2}-1
=\frac{(b-1)\Gamma(\frac{2}{b})}{\Gamma(\frac{1}{b})^2}
\left(1+\frac{\pi}{b-2}\cot\frac{\pi}{b}\right)-1,
\label{bv}
\eeq
implying that the number of particles in a jammed configuration
keeps fluctuating in the thermodynamic limit for all integers $b\ge2$.

For $b=2$, the expression inside the large parentheses in~(\ref{bmu2})
and~(\ref{bv}) becomes $1+\pi^2/4$,
and so the above results coincide with those derived in section~\ref{secde}.
Table~\ref{hib} gives numerical values of the prefactors $\mu_1$ and $\mu_2$
and of the relative variance $V$ for the first few integers $b$.
The lack of self-averaging of the number of particles,
as measured by the size $V$ of its relative fluctuations,
is therefore maximal for $b=2$ and decreases rather fast for higher integer values of $b$.
If the expression~(\ref{bv}) is formally continued to real values of $b$,
the relative variance vanishes both as $b\to1$, according to $V\approx b-1$,
and as $b\to\infty$, according to $V\approx1/b^2$,
and reaches its maximum $V=0.127228\dots$ for $b=1.451602\dots$

\begin{table}[!ht]
\begin{center}
\begin{tabular}{|c|c|c|c|}
\hline
$b$ & $\mu_1$ & $\mu_2$ & $V$\\
\hline
2 & 1.772453\dots & 3.467401\dots & 0.103708\dots\\
3 & 1.339469\dots & 1.905108\dots & 0.061827\dots\\
4 & 1.208536\dots & 1.518872\dots & 0.039924\dots\\
5 & 1.147710\dots & 1.353822\dots & 0.027771\dots\\
6 & 1.113263\dots & 1.264646\dots & 0.020406\dots\\
\hline
\end{tabular}
\caption{Numerical values of the prefactors $\mu_1$ and $\mu_2$
of the first two moments of the total number of particles
(see~(\ref{bmu1}) and~(\ref{bmu2}))
and of the corresponding relative variance $V$ (see~(\ref{bv})),
for the first few integers~$b$.}
\label{hib}
\end{center}
\end{table}

More generally, higher moments of the total number of particles scale as
\beq
\mean{N^k}\approx\mu_k\,L^{k(b-1)/b},
\eeq
where the first two prefactors $\mu_k$ have been derived in~(\ref{bmu1}) and~(\ref{bmu2}).
So, for all integers $b\ge2$,
the probability distribution $P(N,L)$ is expected to scale as
\beq
P(N,L)\approx\frac{\P(\nu)}{L^{(b-1)/b}},
\label{bpro}
\eeq
whenever $N$ and $L$ are both large, with a fixed value of the combination
\beq
\nu=\frac{N}{L^{(b-1)/b}},
\eeq
and the prefactors $\mu_k$ coincide with the moments of the $b$-dependent
limit law $\P(\nu)$:
\beq
\int_0^\infty\nu^k\P(\nu)\,\dd\nu=\mu_k.
\eeq
Matching the scaling law~(\ref{bpro}) with the exact result~(\ref{pblres}) for the probability
of the least dense configurations suggests that $\P(\nu)$ vanishes as a power law, i.e.,
\beq
\P(\nu)\sim\nu^\beta,\qquad\beta=b+\frac{1}{b-1}\qquad(\nu\ll1).
\label{bprosmall}
\eeq
The exponent $\beta$ takes its minimal value $\beta=3$ for $b=2$ (see~(\ref{prosmall})).
Matching~(\ref{bpro}) with the asymptotic expression~(\ref{pbllas}) for the probability
of densest configurations suggests the behavior
\beq
\ln\P(\nu)\sim-\nu^b\ln\nu\qquad(\nu\gg1),
\label{bprolarge}
\eeq
with unit prefactor, irrespective of $b$.

The probability of simultaneous occupation of sequences of $k$ consecutive sites
can also be derived along the same lines.
We thus obtain, for all $k=1,\dots,b$,
\beq
p_{n,n+1,\dots,n+k-1}\approx\int_0^\infty\e^{-nY^b}\rho_k(Y)\,\dd Y,
\label{pseq}
\eeq
where $Y=\max(1-x_n,\dots,1-x_{n+k-1})$ has probability density
$\rho_k(Y)=kY^{k-1}$,
and therefore
\beq
p_{n,n+1,\dots,n+k-1}\approx\frac{\Gamma(\frac{k}{b}+1)}{n^{k/b}}.
\label{pcasy}
\eeq
For $k\ge b+1$,
the ordering of some of the variables $x_n,\dots,x_{n+k-1}$ matters,
and so the simple result~(\ref{pseq}) does not hold any more.

Inserting the expressions~(\ref{pcasy}) into the sum rules~(\ref{sumrules}),
we therefore predict that the mean number of clusters consisting of exactly $k$ sites
grows asymptotically as
\beq
\mean{M_k}\approx\sum_{n=1}^{L+1-k}p_{n,n+1,\dots,n+k-1}
\approx\frac{k\Gamma(\frac{k}{b})}{b-k}\,L^{(b-k)/b}
\eeq
for $k=1,\dots,b-1$.
In the special situation where $k=b$,
the probability $p_{n,n+1,\dots,n+b-1}$ is nothing but
the probability that there is a blockage starting at site~$n$.
This quantity is found to fall off as
\beq
p_{n,n+1,\dots,n+b-1}\approx\frac{1}{n}.
\eeq
This expression generalizes~(\ref{precs}) and~(\ref{ppasy}) to all values of $b$.
As a consequence, the mean number of blockages,
\beq
\mean{M_b}\approx\ln L,
\label{nlog}
\eeq
grows logarithmically with the system size, with unit prefactor, irrespective of $b$.

\section{Summary}
\label{disc}

The theater model introduced in this work is appealing in several regards.
RSA is recognized as being the simplest of all models with fully irreversible dynamics,
whereas the theater model is the simplest local variant of RSA in one dimension,
incorporating directionality and steric constraints as key ingredients.
It is essentially parameter-free,
the only parameters being two integers, the system size $L$
and the number $b$ of particles needed to form a blockage.
The theater model {\it stricto sensu} corresponds to $b=2$.
Last but not least,
as the focus of this study is on the statistics of jammed configurations,
the full stochastic dynamics of the model has been reduced
to questions related to a static random height profile $x_n$,
introduced in~(\ref{xdef}).

The simplest situation where every adsorbed particle is a blockage,
corresponding to $b=1$, was already considered in~\cite{greek}.
It is studied in detail in section~\ref{secun}.
The statistics of jammed configurations can be mapped onto well-known problems
in discrete mathematics,
namely the statistics of records in sequences of independent random variables
and of cycles in random permutations.
The occupations of different sites are statistically independent,
with site $n$ being occupied with probability $p_n=1/n$.
The full distribution of the number $N$ of particles on a system of size $L$ is
related to Stirling numbers of the first kind.
On large systems, this distribution becomes self-averaging,
as all its cumulants grow logarithmically with $L$.
The last three statements still hold in a two-sided variant of the model
where particles may enter from either end of the array,
although the occupations of different sites are not independent any more.

The generic situation where steric interactions are at work,
in the sense that at least two particles have to concur to make a blockage,
has been investigated
in section~\ref{secde} for the theater model {\it stricto sensu} ($b=2$)
and for higher values of~$b$ in section~\ref{sechi}.
Section~\ref{secdeex} contains many exact results of combinatorial nature for finite systems.
The regime of most interest is however that of large systems.
For all integers $b\ge2$, the statistics of jammed configurations exhibits
many common features of interest,
which are surprisingly different both from usual RSA on a homogeneous
substrate and from the case where $b=1$.
The occupations of different sites across the system
exhibit long-range correlations obeying scaling laws.
As a consequence, the total number of particles is not self-averaging.
It rather keeps fluctuating for very large systems,
growing as a subextensive power of the system size, as $N\approx\nu L^{(b-1)/b}$,
where the rescaled variable~$\nu$ has a non-trivial limit law $\P(\nu)$, which depends on~$b$.
A few moments of this limit distribution have been determined,
as well as the form of its decay at small and large~$\nu$.
The mean number of blockages obeys a logarithmic growth law, irrespective of $b$.
It is tempting to infer from this observation that the full statistics of blockages
parallels that of records, which constitute the blockages for $b=1$.
The probability of occurrence of the least dense and densest jammed
configurations has also been scrutinized.
The least dense configurations are those where only the first $b$ sites are occupied.
The probability for the system to end up in a densest, i.e., fully occupied, configuration
has been expressed in terms of the numbers of permutations of $L$ objects
consisting of cycles of length at most $b$.

\ack

It is a pleasure to thank Kirone Mallick for stimulating discussions.
We are also grateful to Sanjay Ramassamy for having made us aware
---after this work was completed---
that he has used the Foata correspondence to establish a bijection
between the densest configurations of the theater model
and permutations consisting of cycles of length at most $b$ (see~\cite{sanjay}).

\appendix

\section{Recursions and asymptotic expansions of occupation probabilities for $b=2$}
\label{app}

\subsection{The single-site occupation probabilities $p_n$}

The explicit expression~(\ref{pires}) of the generating function $\Pi(z)$
implies that the latter quantity obeys the differential equation
\beqa
-(z-1)(z+1)(z+3)\Pi'(z)-4\Pi(z)
&=&2\ln(1-z)
\nonumber\\
&+&(z+1)(2z+3).
\eeqa
As a consequence, the occupation probabilities $p_n$ obey the four-term linear recursion
\beq
(n-2)p_{n-2}+3(n-1)p_{n-1}-(n-4)p_n-3(n+1)p_{n+1}=\frac{2}{n},
\label{prec}
\eeq
for all $n\ge3$.
The above recursion has a special solution
\beq
p_n^{(0)}=\frac{1}{2n},
\eeq
which does not obey the appropriate initial conditions.
Looking for the general solution in the form of an asymptotic expansion in
inverse powers of $n$, we obtain
\beq
p_n=\frac{1}{2n}+\frac{\sqrt{\pi}}{2\sqrt{n}}
\left(1+\frac{1}{16n}+\frac{25}{2^9n^2}-\frac{5}{2^{13}n^3}
-\frac{11781}{2^{19}n^4}+\cdots\right),
\label{pnapp}
\eeq
where the prefactor $\sqrt{\pi}/2$ has been borrowed from the full
solution (see~(\ref{pnasy})),
and therefore
\beqa
\mean{N}
&=&\sqrt{\pi L}\left(1+\frac{3}{16L}-\frac{11}{2^9L^2}
+\frac{69}{2^{13}L^3}-\frac{381}{2^{19}L^4}+\cdots\right)
\nonumber\\
&+&\frac{1}{2}\left(\ln
L+\gamma-3+\frac{1}{2L}-\frac{1}{12L^2}+\frac{1}{120L^4}+\cdots\right).
\eeqa

\subsection{The pair occupation probabilities $p_{n,n+1}$}

The explicit expression~(\ref{pi1res}) of the generating function $\Pi_1(z)$
implies that the latter quantity obeys the differential equation
\beqa
-z(z+1)(z+3)\Pi_1'(z)-(z-3)\Pi_1(z)
&=&2\ln(1-z)
\nonumber\\
&+&2z(z+2).
\eeqa
As a consequence, the pair occupation probabilities $p_{n,n+1}$ obey the
three-term linear recursion
\beq
(n-2)p_{n-2,n-1}+(2n-1)p_{n-1,n}-3(n+1)p_{n,n+1}=-\frac{2}{n},
\label{pprec}
\eeq
for all $n\ge3$.
The above recursion has a special solution
\beq
p_{n,n+1}^{(0)}=p_n^{(0)}+p_{n+1}^{(0)}=\frac{2n+1}{2n(n+1)},
\eeq
which does not obey the appropriate initial conditions.
Looking for the general solution in the form of an asymptotic expansion in
inverse powers of $n$, we obtain
\beqa
p_{n,n+1}&=&\frac{1}{n}\left(1-\frac{1}{2n}+\frac{1}{2n^2}
-\frac{1}{2n^3}+\frac{1}{2n^4}+\cdots\right)
\nonumber\\
&+&
\frac{\sqrt{\pi}}{2n\sqrt{n}}
\left(1-\frac{15}{16n}+\frac{505}{2^9n^2}-\frac{8085}{2^{13}n^3}
+\frac{505659}{2^{19}n^4}+\cdots\right),
\label{ppnapp}
\eeqa
where the first line is the expansion of the special solution,
and the prefactor of the second line has been borrowed from the full solution
(see~(\ref{ppasy})).

\section*{References}

\bibliography{paper.bib}

\end{document}